\documentclass[journal]{IEEEtran}
\ifCLASSINFOpdf
\else
\fi
\usepackage{soul, color, xcolor}
\sethlcolor{yellow}

\soulregister{\cite}7
\soulregister{\citep}7
\soulregister{\citet}7
\soulregister{\ref}7
\soulregister{\pageref}7
\usepackage{bm}
\usepackage{amsfonts}
\usepackage{amsmath,amsfonts,amssymb,epsfig,cite,url,multirow,booktabs}
\usepackage{afterpage,mathrsfs,algorithm,algorithmic}
\usepackage{amsthm}
\usepackage{bbding}

\graphicspath{{./pic/}}
\usepackage{makecell}
\usepackage{graphicx}
\usepackage{float}
\usepackage{subfig}
\usepackage{caption}
\usepackage{array, booktabs}

\begin{document}
%
\title{UAV Individual Identification via Distilled RF Fingerprints-Based LLM in ISAC Networks }
%

%

\author{Haolin Zheng,~Ning Gao,~\IEEEmembership{Member,~IEEE,}~Donghong Cai,~\IEEEmembership{Member,~IEEE,}~Shi Jin,~\IEEEmembership{Fellow,~IEEE,}\\and~Michail Matthaiou,~\IEEEmembership{Fellow,~IEEE}

\thanks{This work was supported in part by the National Key Research and Development Program of China under Grant 2024YFE0200700, in part by National Science Foundation of China (NSFC) under Grant 62371131, and in part by the European Research Council (ERC) under the European Union’s Horizon 2020 research and innovation programme under Grant 101001331.}
\thanks{H. Zheng and N. Gao are with the School of Cyber Science and Engineering, Southeast University, Nanjing 210096, China (e-mail: 3150875990@qq.com; ninggao@seu.edu.cn).}

\thanks{D. Cai is with the College of Cyber Security, Jinan University, Guangzhou 510632, China (e-mail: dhcai@jnu.edu.cn).}
\thanks{S. Jin is with the National Mobile Communications
Research Laboratory, Southeast University, Nanjing 210096, China (e-mail: jinshi@seu.edu.cn).}
\thanks{
M. Matthaiou is with the Centre for Wireless Innovation (CWI), Queen’s University Belfast, Belfast BT3 9DT, U.K. (e-mail: m.matthaiou@qub.ac.uk).}
}
\markboth{}%
{Shell \MakeLowercase{\textit{et al.}}: Bare Demo of IEEEtran.cls for IEEE Journals}
%



\maketitle

\begin{abstract}
Unmanned aerial vehicle (UAV) individual (ID) identification is a critical security surveillance strategy in low-altitude integrated sensing and communication (ISAC) networks. In this paper, we propose a novel dynamic knowledge distillation (KD)-enabled wireless radio frequency fingerprint large language model (RFF-LLM) framework for UAV ID identification. First, we propose an RFF-LLM framework based on the modified GPT-2 model to improve the identification accuracy in complex outdoor environments. Then, considering the parameter overhead of the RFF-LLM, we design a dynamic KD strategy to compress the model. Specifically, the proximal policy optimization (PPO) algorithm is employed to dynamically adjust the distillation temperature, overcoming the local optimum dilemma inherent in static KD. As a next step, the knowledge of the RFF-LLM is adequately transferred to the lightweight Lite-HRNet model. Finally, our experiments are conducted based on the self-built drone RFF dataset of Release one, namely DRFF-R1, by collecting the I/Q signals of 20 commercial UAVs in channel 149. The experiment results show that the proposed framework achieves 98.38\% ID identification accuracy with merely 0.15 million parameters and 2.74 ms response time, which outperforms the benchmarks.
\end{abstract}


%
\IEEEpeerreviewmaketitle
\section{Introduction}
\IEEEPARstart{W}{ith} the evolution of 5G-Advanced (5G-A) integrated sensing and communication (ISAC) technology, base stations can implement precise positioning of unmanned aerial vehicles (UAVs) through multi-dimensional sensing information\cite{10004900}. However, traditional UAV surveillance can only perceive the UAV physical state, but cannot recognize the UAV identity. In particular, most of the consumer-grade UAVs still have not been registered online. Thus, during malicious incidents, such as illegal reconnaissance, privacy infringement, or terrorist attacks, the traditional surveillance struggles to authenticate the UAV, making it difficult to establish users accountable\cite{10463115}. As a result, the traditional physical surveillance methods are not effective in complex non-cooperative low-altitude scenarios.

Radio-frequency (RF) fingerprint originates from hardware imperfections during the device manufacturing processes, representing inherent physical layer characteristics, such as carrier frequency offset and in-phase/quadrature (I/Q) imbalance\cite{ARXIV25}. Compared to traditional authentication methods, like Internet Protocol (IP) addresses, media access control (MAC) addresses, etc., radio frequency fingerprint (RFF) identification establishes a security authentication mechanism beyond the bit-level, whose physical unclonable properties make it exceptionally resistant to malicious imitation\cite{GU2024110115}. This technology has achieved breakthroughs in underpinning the desirable Internet-of-Things (IoT) device authentication. Notably, RFF identification demonstrates substantial potential in UAV \underline{i}n\underline{d}ividual (ID) identification. However, most of the current works primarily focus on UAV type classification \cite{10737118,fu2025towards}, but pay less attention to UAV ID identification.

Recently, Li \textit{et al.} achieved an 98.91\% accuracy through time-domain instantaneous amplitude statistics and fractal dimension feature extraction, validating the efficacy of using RFF for UAV ID identification, but the feature engineering introduces a large overhead\cite{li2024novel}. Several recent works have studied the automatic feature extraction for signal recognition based on deep learning, and have demonstrated significant advances in recognition accuracy and inference speed \cite{9049161,9200788,cai2024toward}. Nevertheless, the current works on UAV ID identification are relatively scarce, which is due to the challenge of the low ID identification efficiency and high cost in complex outdoor environments, and the lack of enough UAV ID identification datasets\cite{kunze2024radio,cai2024toward,li2024novel}. Against this background, the recent advancements in Transformer-based large language models (LLMs) have illuminated their exceptional performance across various language processing tasks. Very recently, the LLMs have also found significant applications in wireless communications space\cite{10531073}. For the first time, Gao \textit{et al.} migrated knowledge from a pre-trained BERT model to the LightRFFI model, maintaining a remarkable 97.52\% identification accuracy\cite{gao2025rff}. While the static knowledge distillation (KD) limits the LLM efficiency, the fixed distillation temperature configuration often trap lightweight models in local optima\cite{choudhary2020comprehensive}. Motivated by the above mentioned challenges and opportunities, in this paper, we propose a novel dynamic KD enabled wireless radio frequency fingerprint large language model (RFF-LLM) framework to improve the ID identification efficiency of UAVs. The main contributions of this paper are listed as follows:\vspace{-0.5em}
\begin{itemize}
\item By extracting high-dimensional temporal features and long-range dependencies from the time-frequency spectra, we propose a dynamic KD-enabled wireless RFF-LLM framework based on the modified GPT-2, which improves the ID identification accuracy in complex outdoor environments.
\item Considering the deployment costs and efficiency, we adequately transfer the knowledge of RFF-LLM to the lightweight Lite-HRNet model through the proximal policy optimization (PPO)-based dynamic distillation, which overcomes the local optima dilemma of static distillation.
\item We develop a prototype platform to collect I/Q signals of 20 UAVs covering 7 types, constructing and publicly releasing the UAV RFF dataset. The results show that the dynamically distilled Lite-HRNet model converges to global optima within 61 iterations, achieving an impressive 98.38\% UAV ID identification accuracy with merely 0.15 million parameters and 2.74 ms response time.
\end{itemize}

\section{System Model}
\subsection{Signal Reception}
We consider a single antenna for UAV, with line-of-sight (LoS) and non-line-of-sight (NLoS) links in outdoor environments. The channel is modeled using the Ricean fading model
\begin{equation}
H(t) = \sqrt{\frac{K}{K+1}}h(t)+ \sqrt{\frac{1}{K+1}}\tilde{h}(t),
\end{equation}
where \( K \) represents the Ricean factor, \(h(t)\) represents the direct link component and \(\tilde{h}(t) \sim \mathcal{CN}(0,1)\)  represents the equivalent scattered component. In this case, the received signal \(s(t)\) is represented as
\begin{equation}
	s(t) = H(t) \otimes b(t) + n(t),
\end{equation}
where \(n(t) \sim \mathcal{CN}(0,\sigma^2)\) is  the environmental noise, the symbol \( \otimes \) denotes the convolution operation, and \(b(t)\) is the baseband I/Q transmit signal. Considering the hovering state of the UAV and that the receiver is stationary and unique, there are no Doppler effect and influence of the RFF of the receiver. The ideal baseband I/Q signal model is
\begin{equation}
	b(t) = b_I(t) + jb_Q(t),
\end{equation}
where \( b_I(t),b_Q(t)\) denote the time-varying baseband components. However, accounting for the hardware imperfections of the transmitter, this signal model can be revised to
\begin{equation}
	b(t) = \alpha\cos(2\pi f_0 t + \varphi)b_I(t) - j\sin(2\pi f_0 t)b_Q(t),
\end{equation}
where \( f_0 \) is the carrier frequency, \( \alpha \in (0,1] \) denotes the transmitter gain imbalance, whereas  \( \varphi \in (0,2\pi] \) represents the phase imbalance.

At the receiver side, the received signal \(s(t)\) undergoes discrete sampling upon reception, resulting in the final discrete signal \(S[n]\), which is given by
\begin{equation}
	S[n] = s(nT_s) = I[n] + jQ[n],\quad n = 0,1,...,N-1,
\end{equation}
where \( T_s \) denotes the sampling interval, with \(I[n]=\Re[s(nT_s)]\quad \text{and} \quad Q[n]=\Im[s(nT_s)]\) represents the in-phase and quadrature components, respectively. The discrete sampling signal \(S[n]\) is transformed into the raw data \(x_l\), where \(l\in L\) denotes the true label, and \(L\) denotes the number of labels.
\begin{figure}[!ht]
	\centering
	\includegraphics[width=8.8cm]{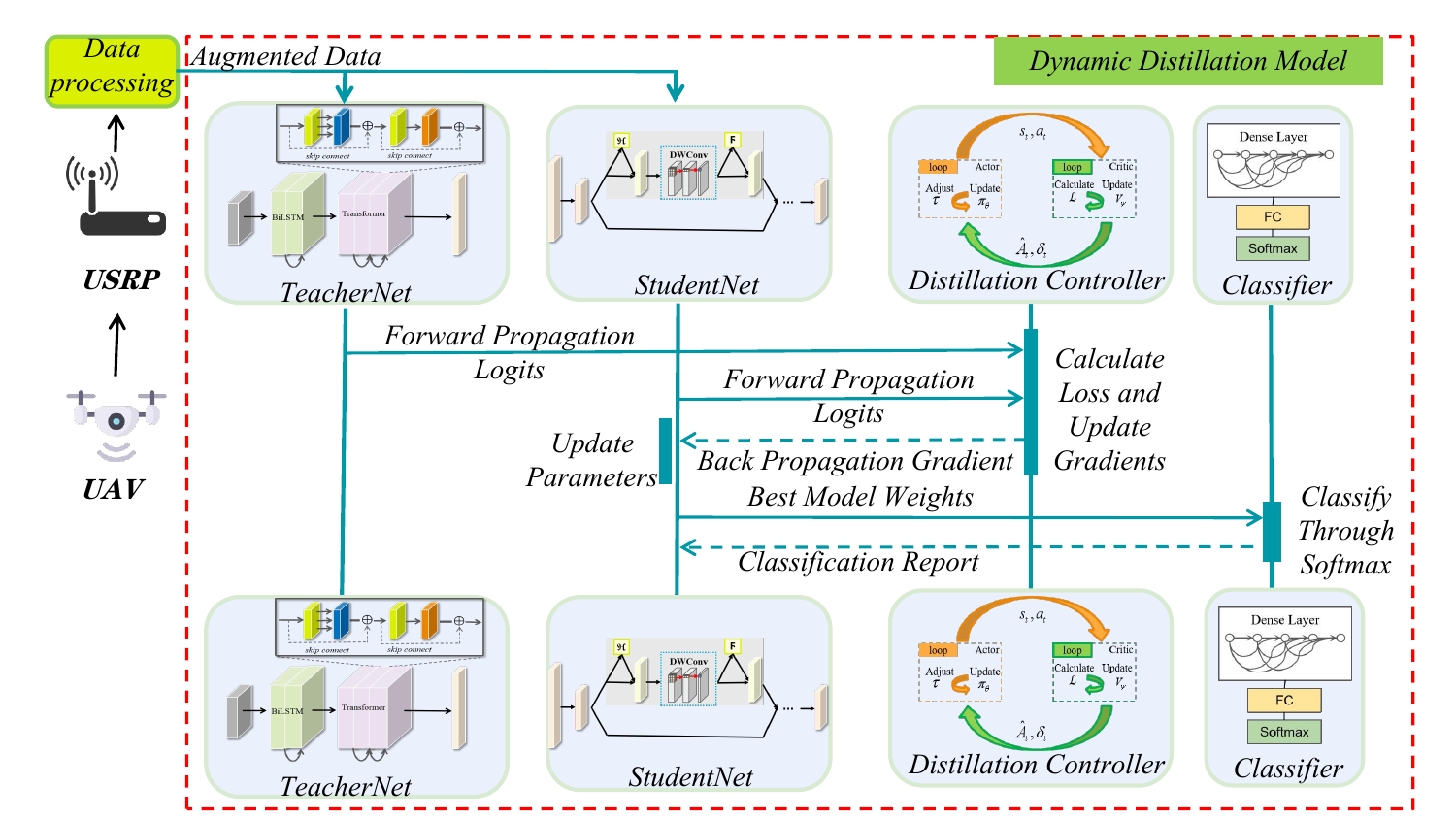}
	\caption{The overview of the proposed framework.}
	\label{Fig:frame}
\end{figure}
The augmented signal samples \(\tilde{x}_l=\textit{Augment}(x_l)\)  constitute the sample space \(\mathbf{X}=\{\tilde{x}_l|l\in1,\ldots, L\}\), and then we can construct the dataset \( \mathcal{D} = (\tilde{x}_l, l) \substack{\tilde{x}_l \in \mathbf{X} \\l \in L} \).

\subsection{Problem and Framework Formulation}

The problem of UAV ID identification can be formulated as a multi-class classification task based on the augmented signal samples \(\tilde{x}_l\). The model learns the mapping from \(\tilde{x}_l\) to \(l\)  by minimizing the cross-entropy (CE) loss \(\mathcal{L}_{\text{CE}}\)  between the predicted label \(\hat{l}\)  and the true label \(l\) to achieve UAV ID identification. The optimization objective is expressed as
\begin{equation}
	\min_{\eta} \mathbb{E}_{(\tilde{x},l)\sim\mathcal{D}}\left[\mathcal{L}_{CE}(\hat{l},l,\tilde{x},\eta)\right],
\end{equation}
where \(\eta\) represents the model parameters and \(\mathbb{E}[\cdot]\) represents the mathematical expectation. As illustrated in Fig. \ref{Fig:frame}, the proposed dynamic KD-enabled wireless RFF-LLM framework comprises four key components: teacher network RFF-LLM, student network Lite-HRNet, distillation controller and classifier. The dynamic distillation employs a dual-phase optimization strategy: Initially, the teacher network parameters \(\mathbf{W}_T\) are fixed, generating logits \(k_T=f_{\text{teacher}}(\mathbf{X};\mathbf{W}_T)\colon\mathbb{R}^d \rightarrow \mathbb{R}^c\). The student network produces logits \(k_S=f_{\text{student}}(\mathbf{X};\mathbf{W}_S)\colon\mathbb{R}^d \rightarrow \mathbb{R}^c\), while \(k_T,k_S\) are forward propagated to the distillation controller to compute the CE loss \(\mathcal{L}_{\text{CE}}\) and Kullback-Leibler (KL) divergence loss \(\mathcal{L}_{\text{KL}}\). Thus, the total loss function can be denoted as $\mathcal{L}_{\text{TOTAL}} = (1-\beta)\mathcal{L}_{\text{CE}}+\beta\mathcal{L}_{\text{KL}}$, with distillation weight \(\beta\), which seeks to update the parameter $\mathbf{W}_S$ of the student network via back propagating gradients. In particular, the PPO actor network  dynamically adjusts the distillation temperature \(\tau\) based on the state information \(s_t\) and the reward function \(R\). The joint optimization problem is formalized as
\begin{gather}
	\min_{\mathbf{W}_S} \mathbb{E}_{(\tilde{x},l)\sim\mathcal{D}} \left[\mathcal{L}_{\text{TOTAL}}(\tilde{x},l,\tau;\mathbf{W}_S,\mathbf{W}_T)\right], \\
	\max_{\theta} \mathbb{E}_{s_t\sim\pi_\theta}\left[R(s_t;\tau)\right].
\end{gather}

The optimized student parameters \(\mathbf{W}_S\) are delivered to the classifier, which executes the final classification and returns the classification result.
\section{Methodology}
\subsection{RFF-LLM Framework}
The GPT-2 model utilizes absolute positional encoding to capture the positional information in sequences. However, when processing time-frequency spectra generated by short-time Fourier transform (STFT), absolute positional encoding struggles to model the complex temporal dependencies effectively, resulting in limited identification accuracy even after feature extraction through Transformer layers. Moreover, the original GPT-2 framework with over 12 stacked Transformer layers is easy to overfitting when training data is insufficient. Thus, the architecture of the original GPT-2 model is modified to adapt to the wireless I/Q signal, which is detailed as follows:

\textbf{Model Architecture:} The STFT time-frequency spectra first pass through two stacked BiLSTM layers, generating temporal feature tensors, and then processed by three Transformer layers for long-range dependency modeling, enhanced by residual connections. Finally, classification is achieved through fully connected layers and the number of classes.

\textbf{Embedding:} We employ the BiLSTM network as the core temporal feature extraction module. The forward and backward hidden states are obtained through the gated computation processes defined in (\ref{eq:forward}), (\ref{eq:backward}), then fused via the concatenation operation in (\ref{eq:concat}) to output the features. This design replaces the absolute position encoding in the original GPT-2 architecture, enabling adaptive generation of dynamic position-aware deep feature representations through the LSTM's temporal modeling capability, which is more suitable for spatio-temporal feature learning in non-stationary signals.	The procedure can be represented as
\begin{align}
	\label{eq:forward}
	z \substack{f \\ t} &= \textit{LSTM}(x_t, z \substack{f \\ t+1}), \\
	\label{eq:backward}
	z \substack{b \\ t} &= \textit{LSTM}(x_t, z \substack{b \\ t+1}), \\
	\label{eq:concat}
	\mathcal{Z}_t &= [z \substack{f \\ t} || z \substack{b \\ t}],
\end{align}
where \(z \substack{f \\ t},z \substack{b \\ t}\) denote the forward and backward hidden states at time step \(t\), respectively,  \(x_t\) represents the data feature, while \(\mathcal{Z}_t\)  indicates the concatenated fusion features.

\textbf{Classifier:} To design the classifier, we employ a global average pooling operation along the temporal dimension to aggregate features from all hidden states, replacing the original output layer design that projects the last hidden state to the vocabulary space. For regularization strategies, we integrate the weight regularization of each fully connected layer with stochastic dropout, constructing a multi-level regularization system, which is mathematically implemented as
\begin{equation}
\mathcal{L}_{\text{reg}} = \lambda \sum_{l=1}^{L} \| \mathbf{W}_l \|_F^2,
\end{equation}
where \(\mathbf{W}_l\) represents the weight matrix of the \(l\text{-th}\) fully connected layer, \(L\) indicates the total number of fully connected layers in the classification head, \(||\cdot||_F\) represents the Frobenius norm, while \(\lambda\) is the regularization coefficient.

\subsection{Dynamic Distillation With PPO}

The temperature depends only on the current state-action pair, therefore satisfying the Markov property, which can be modeled as an Markov decision process (MDP) with the tuple \((\boldsymbol{S}, \boldsymbol{A}, \boldsymbol{P}, \boldsymbol{R}, \gamma)\), where \(\boldsymbol{P}\) is implicitly conducted through an experience replay buffer mechanism, while \(\gamma\) is the discount factor with a fixed value. Specifically, the three tuple \((\boldsymbol{S}, \boldsymbol{A}, \boldsymbol{R})\) is given by

\noindent \textbf{State Space \(\boldsymbol{S}\)}: We define the state information as $
	s_t = \left( \mu_{\text{acc}}^{k}, \sigma_{\text{acc}}^{k}, \Delta_{\text{acc}}^{k}, \mu_{\text{KL}}^{k}, \sigma_{\text{KL}}^{k}, \Delta_{\text{KL}}^{k}, \zeta \right)$,
where \(\mu_{\text{acc}}^{k}, \sigma_{\text{acc}}^{k},\Delta_{\text{acc}}^{k}\) denote the \(k\)-step moving average, standard deviation, and linear change rate of accuracy,  \(\mu_{\text{KL}}^{k}, \sigma_{\text{KL}}^{k}, \Delta_{\text{KL}}^{k}\) respectively denote the \(k\)-step moving average, standard deviation, and linear change rate of KL-divergence, while \(\zeta\)   is the training progress.

\noindent \textbf{Action Space \(\boldsymbol{A}\)}: Action is defined as \(a_t\), where the actor network \(\pi_\theta\) outputs the mean \(\mu_\theta\) and variance \(\sigma_\theta\) of a Gaussian distribution. Using (\ref{eq:sample}) for action sampling, the action is converted to the action space \([\tau_{\min},\tau_{\max}]\) through a linear transformation, which is given by
\begin{align}
	\label{eq:sample}
	a_t &=\mu_\theta(s_t)+\varepsilon_t, \varepsilon_t \sim \mathcal{N}(0,\sigma_{\theta}^{2}), \\
	\label{eq:convert}
	\tau_t &=\tau_{\min}+a_t(\tau_{\max}-\tau_{\min}).
\end{align}
\noindent \textbf{Reward \(\boldsymbol{R}\)}: The reward function can be designed as
\begin{align}
R_t &= w_1(\xi_t - \xi_{\text{base}}) + w_2 \log[1 + 10(k_t - k_{\text{target}})^2] \nonumber \\
&\quad \quad \quad \quad \quad + w_3 |\tau_t - \tau_{t-1}|^\rho,
\end{align}
where  \(\xi_t\) is the batch-wise accuracy, \(\xi_{base}\) is an accuracy threshold, \(k_t\) is the KL divergence between two networks, and \(k_{\text{target}}\) is a target divergence, \(\rho\) is a custom exponent, while \(w_i,i=1,2,3\) are custom weights.

Considering the continuous and the dynamic nature of temperature adjustment in KD and the training overhead, the model-free PPO algorithm is adopted for its ability to achieve stable policy improvement while keeping computational complexity low. The formulation of PPO is given by \cite{schulman2017proximal}
\begin{gather}
	\label{eq:clip}
	L_{t}^{\text{CLIP}}(\theta)=\mathbb{E}_t[\min(r_t\hat{A}_t,\mathrm{clip}(r_t(\theta),1-\kappa,1+\kappa)\hat{A}_t], \\
	\label{eq:ratio}
	r_t(\theta)=\frac{\pi_\theta(a_t|s_t)}{\pi_{\theta_\text{old}}(a_t|s_t)}, \\
	\label{eq:GAE}
	\hat{A}_t=\sum_{k=0}\gamma^k\delta_{t+k},\delta_t=R_t+\gamma V_\psi(s_{t+1})- V_\psi(s_{t}),
\end{gather}
where \( r_t(\theta) \) is the policy ratio, and \(\hat{A}_t\) is the advantage estimation via generalized advantage estimator (GAE). This formulation aligns with the policy gradient theorem in \cite{sutton1999policy} by (\ref{eq:gradient}), thereby providing gradient update directions
\begin{equation}
	\label{eq:gradient}
	\nabla_\theta J(\pi_\theta)=\mathbb{E}_\pi[\gamma^t\nabla_\theta\log\pi_\theta(A_t|S_t) Q_\pi(A_t|S_t)].
\end{equation}

The process is summarized as \textbf{Algorithm \ref{alg:critic_update}}. For the dynamic temperature adjustment, we set the accuracy threshold as \(A_{\text{base}}\) and reduce the temperature fluctuations by limiting the exploration rate \( \sigma_\theta\), which is detailed in \textbf{Algorithm \ref{alg:temp_adjust}}.

\section{Experimental Results}

\subsection{Experimental Setup}
\begin{figure}[!ht]
	\centering
	\includegraphics[width=8.5cm]{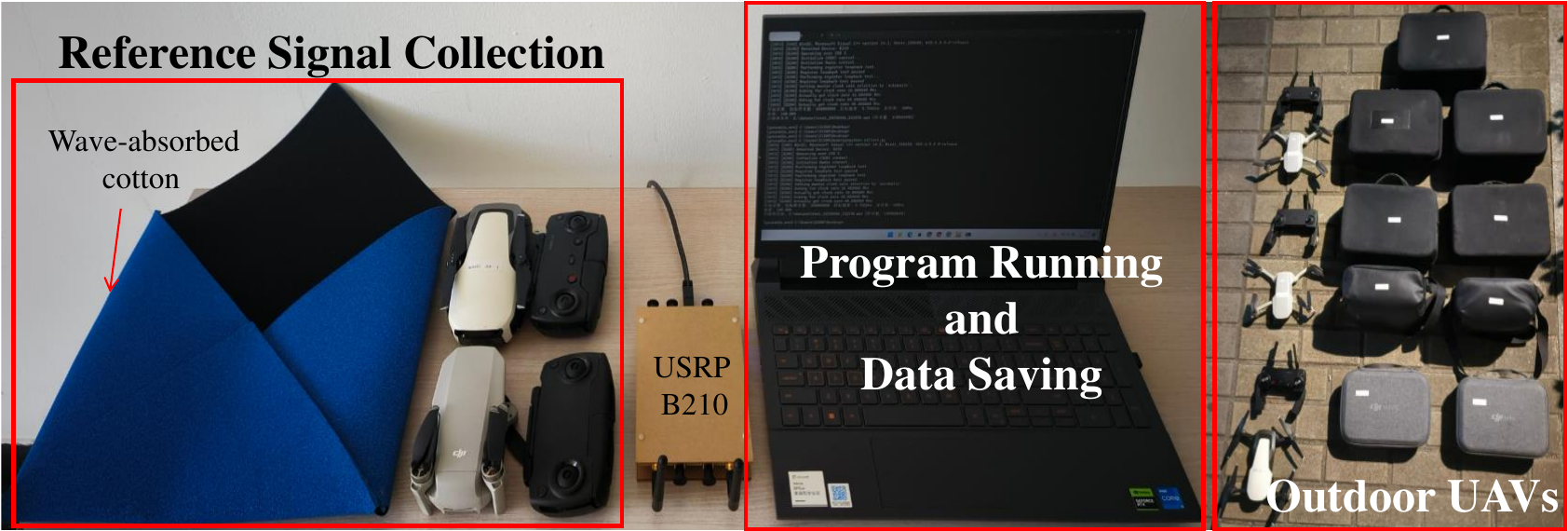}
	\caption{Schematic diagram of the developed prototype system.}
	\label{Fig:platform}
\end{figure}

All experiments are conducted on server clusters equipped with NVIDIA Tesla V100 (32 GB) GPUs, Intel Xeon E5-2698 CPUs, and 128 GB RAM. Our network model development utilizes TensorFlow 2.2 framework. The experimental dataset is self-built with real-world experiments. Specifically,
as shown in Fig. \ref{Fig:platform}, a USRP B210 is deployed to capture the uplink I/Q signals from 20 commercial drones of 7 models at multiple altitudes (10 m, 30 m, 50 m, 70 m, 90 m) in outdoor environments, with the interference-free reference signals at 0 m acquired under wave-absorbing cotton. This real-world dataset is publicly available on GitHub.\footnote{https://github.com/zz-zz-cyber/DRFF-R1} The data of multiple altitudes are used and the time-frequency diagrams obtained through STFT are segmented into 20,000 samples, including the training, validation, and the test sets at a 6:2:2 ratio.

 \begin{figure}[!ht]
	\centering
	\includegraphics[width=8cm]{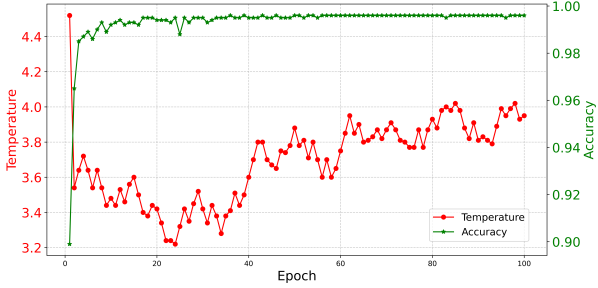}
	\caption{Parameter fluctuations in dynamic distillation.}
	\label{Fig:tem_acc}
\end{figure}

\subsection{Dynamic Distillation Feasible Analysis}
To validate the effectiveness of the proposed strategy, we compare the performance of Lite-HRNet under three configurations, including the no KD (NKD), the fixed temperature distillation ($\tau=2/4/6/8$), and the dynamic distillation. The results show that the dynamic distillation achieves a test accuracy of 98.38\%, outperforming the best fixed temperature setting, i.e., \(\tau=4\) with accuracy 96.62\% and baseline, i.e., 95.28\%. As illustrated in Fig. \ref{Fig:tem_acc}, during the initial training stage, temperature gradually decreases, enabling the student network to effectively extract fine-grained feature knowledge from the teacher network, driving the validation accuracy to rapidly improve from 89.9\% to 99.00\% within the first 5 epochs. By exploring the intermediate stage, i.e., epoch 5-60, the distillation temperature converges to a stable range of \(\tau=3.90\pm0.10\), while maintaining the stable validation accuracy at \(99.60\%\pm 0.10\%\) after 61 training epochs. This convergence characteristic of temperature and identification accuracy confirms that the dynamic strategy effectively overcomes the local optimum dilemma in static distillation.

\begin{figure*}[!ht]
	\centering
	\includegraphics[width=17.4cm]{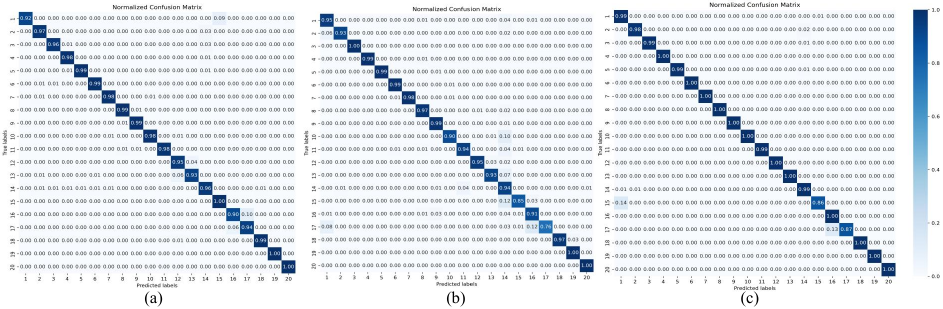}
	\caption{Confusion matrix: (a) RFF-LLM, \textit{Acc}=97.17\%  (b) Lite-HRNet-NKD, \textit{Acc}=95.28\%  (c) Lite-HRNet-KD, \textit{Acc}=98.38\%.}
	\label{Fig:matrix}
	\vspace{-0.1em}
\end{figure*}

To demonstrate the UAV ID identification performance of the Lite-HRNet-KD, we conduct a comparative analysis of confusion matrices across different models. The results in Fig. \ref{Fig:matrix} reveal that the teacher network RFF-LLM maintains high accuracy across all classes, whereas Lite-HRNet-NKD exhibits performance degradation in specific classes, e.g., in Class 17 the accuracy drops to 0.76. Through dynamic KD, Lite-HRNet-KD synergistically integrates the teacher network's feature space distribution priors with its own feature representations, achieving a 11\% relative improvement in weak classes, while enhancing high-accuracy classes, e.g., Class 1 improves from 0.95 to 0.99. Interestingly, it attains a mean accuracy of 98.38\%, outperforming the teacher network by 1.23\%.
\subsection{Comparison of Different Models}
\begin{figure}[h]
	\centering
	\subfloat[]{ \includegraphics[width=0.48\linewidth]{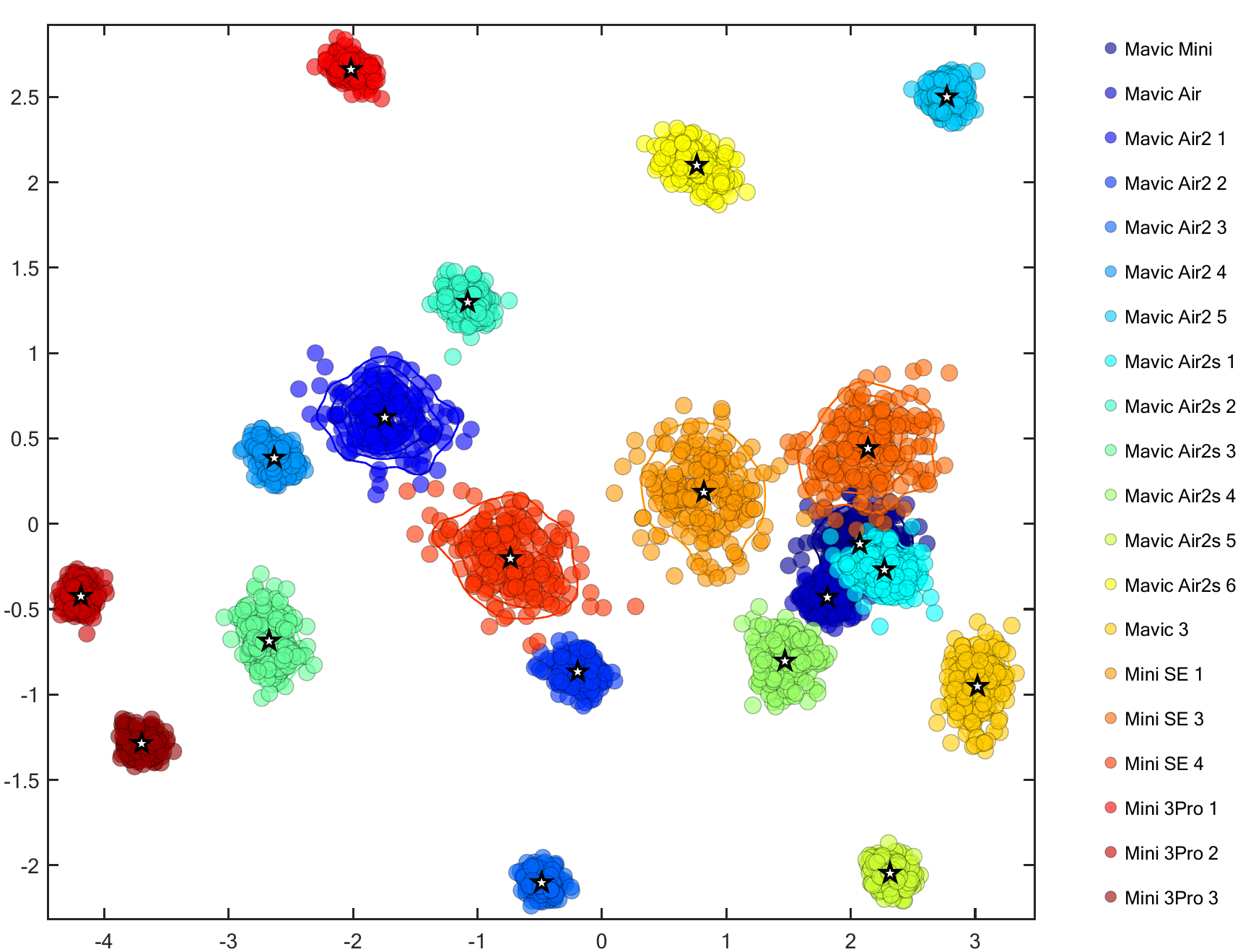}}%
	\hfil
	\subfloat[]{
		\includegraphics[width=0.48\linewidth]{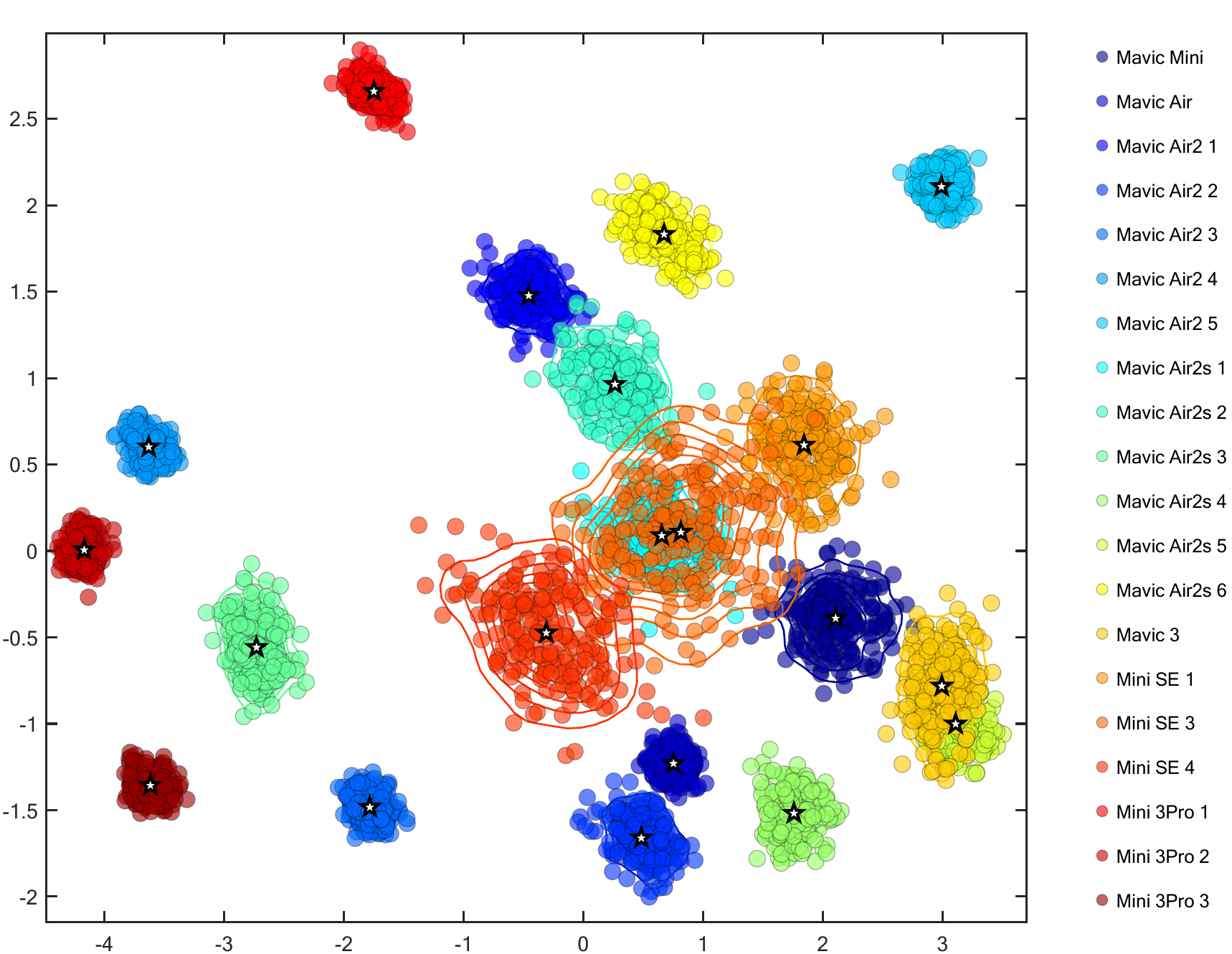}}%
	\hfil
	\subfloat[]{
	\includegraphics[width=0.48\linewidth]{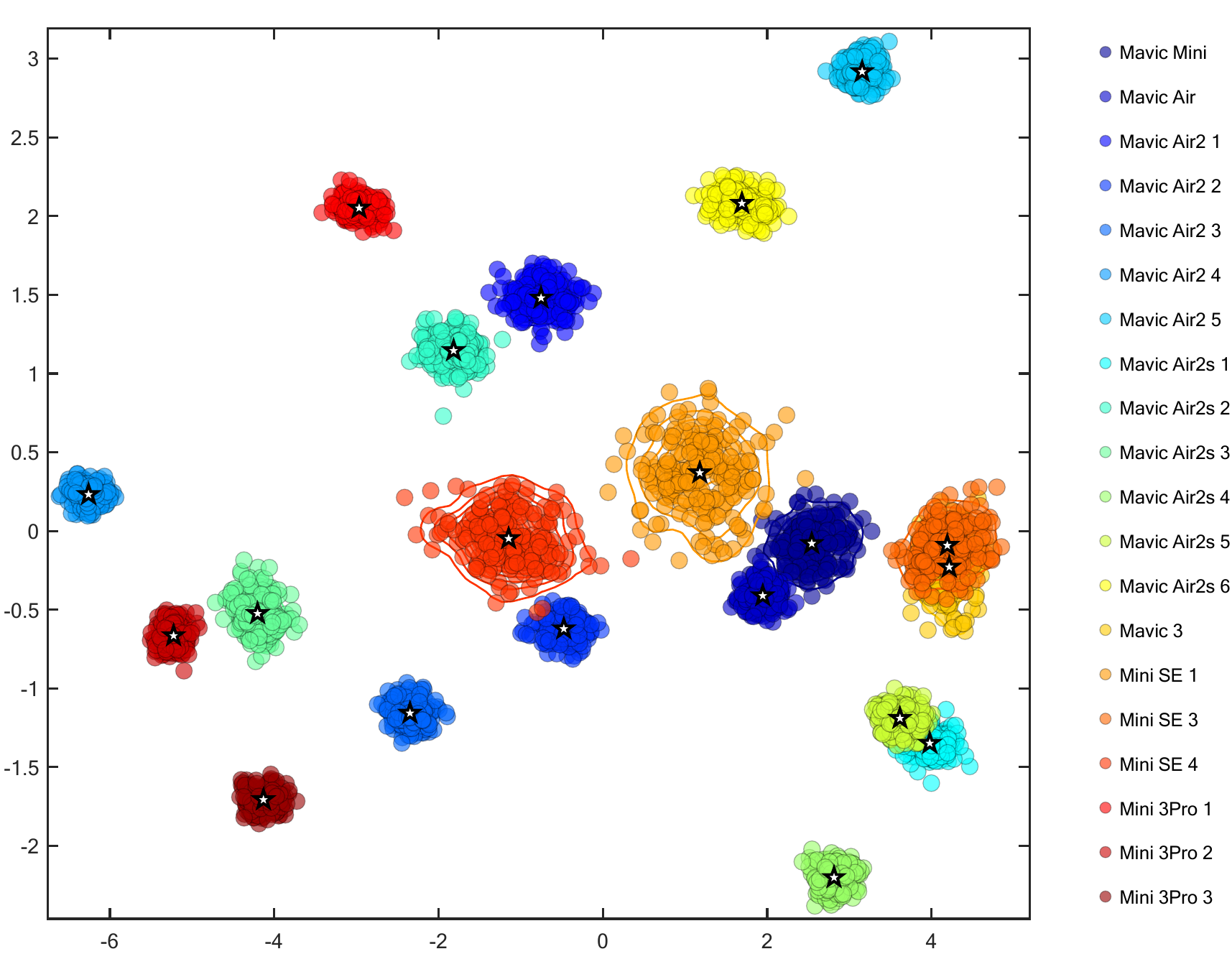}}%
	\hfil
	\subfloat[]{ \includegraphics[width=0.48\linewidth]{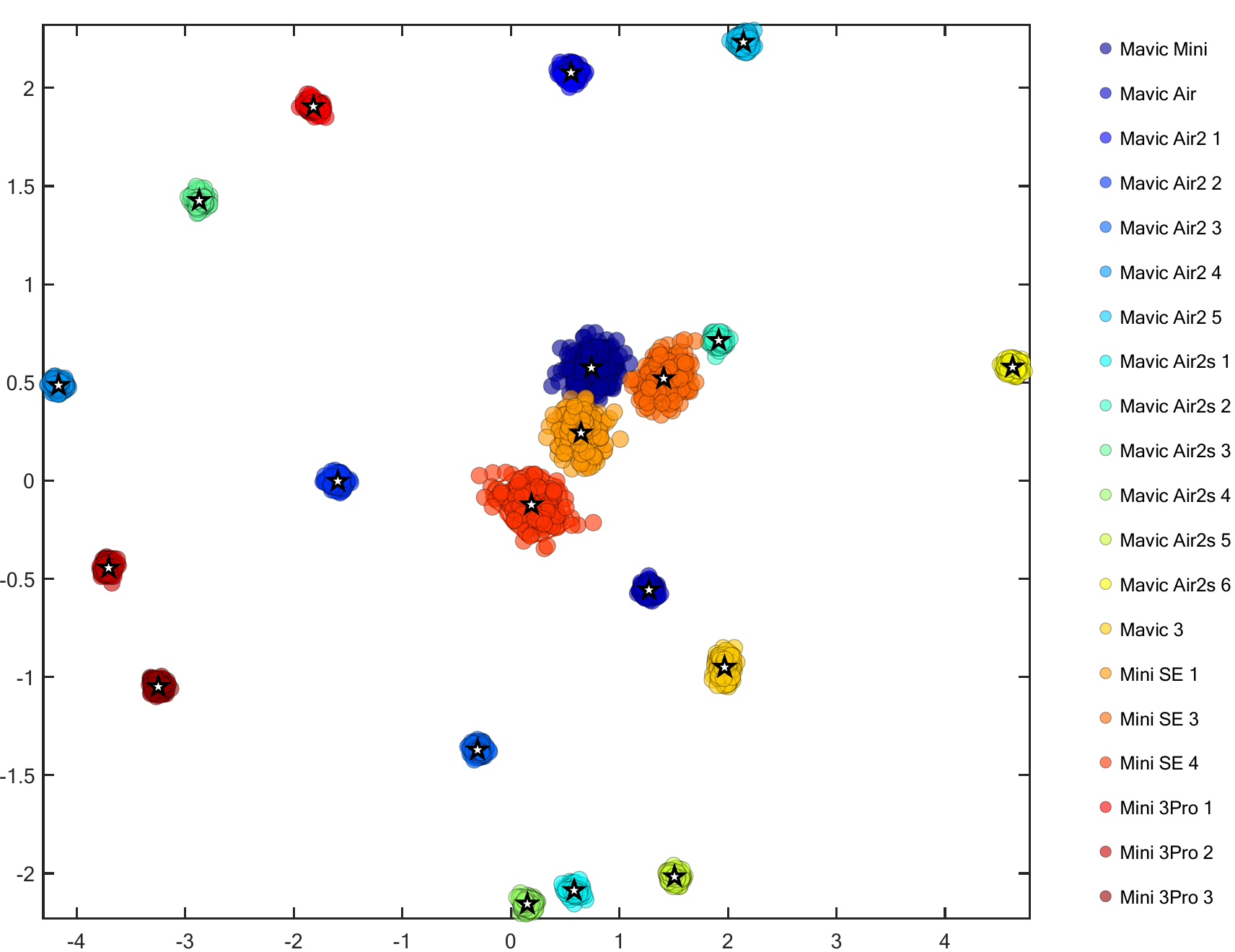}}%
	\captionsetup{justification=raggedright, singlelinecheck=false}
	\caption{Feature visualization: (a) ResNet (b) LMSC (c) ShuffleNet-v2 (d) Lite-HRNet-KD.}
	\label{Fig:visualization}
	\vspace{-1em}
\end{figure}

We now perform a comparison between the Lite-HRNet-KD and the recent state-of-the-art methods, including ResNet \cite{cai2024efficient}, LMSC \cite{cai2024toward}, and ShuffleNet-v2 \cite{ma2018shufflenet}. As shown in Table \ref{tab:model_comparison}, Lite-HRNet-KD achieves a 98.38\% accuracy with the lowest number of parameters, significantly outperforming the three benchmarks. On the same test set, the Lite-HRNet-KD attains 2.74 ms inference latency with lowest RAM usage, showing 24.13\% usage decrease over the suboptimal ShuffleNet-v2. The results validate the real-time security surveillance performance of the  proposed network on resource-constrained devices. The feature space visualization in Fig. \ref{Fig:visualization} further reveals that through dynamic distillation, Lite-HRNet-KD exhibits tighter intra-class clustering and further between-class distance after principal component analysis (PCA) dimensionality reduction. This phenomenon is attributed to the adequate transfer of high-order feature knowledge from the RFF-LLM to the Lite-HRNet via dynamic distillation.

\begin{algorithm}[H]
	\caption{Actor Network Policy Update}
	\label{alg:critic_update}
	\begin{algorithmic}[1]
		\REQUIRE Actor Network \(\pi_\theta\), Critic Network \(V_\psi\).
		\ENSURE  Actor Network Parameter \(\pi_{\theta'}\).
		
		\STATE \textbf{function} policy\_update:
		\STATE Collect trajectory \((a_t, s_t, r_t, \log (\pi_{\theta_{\text{old}}}(a_t|s_t)))\);
		\STATE Compute advantage estimates \(\hat{A}_t\) using GAE;
		\STATE Normalize advantage \(\hat{A}_t \leftarrow \frac{\hat{A}_t-\mu}{\sigma}\) with batch-wise advantage mean \(\mu\) and variance \(\sigma\);
		\STATE Compute probability ratio \(r_t(\theta)\);
		\STATE Calculate clipped objective \(L_t^{\text{CLIP}}(\theta)\);
		\STATE Add entropy regularization with entropy coefficient \(\omega\)
		\[
		L_t^{\text{TOTAL}} = L_t^{\text{CLIP}} + \omega \mathbb{E}_t\left[\mathcal{H}(\pi_\theta(\cdot|s_t))\right];
		\]
		\STATE Update \(\theta' \leftarrow \theta + \nu \nabla_\theta L_t^{\text{TOTAL}}\) with learning rate $\nu$.
	\end{algorithmic}
\end{algorithm}
\begin{algorithm}[H]
	\caption{Dynamic Temperature Adjustment}
	\label{alg:temp_adjust}
	\begin{algorithmic}[1]
		\REQUIRE Temperature range \([\tau_{\min}, \tau_{\max}]\), accuracy threshold \(A_{\text{base}}\).
		\ENSURE Optimized student model weights $\mathbf{W}_S$.
		\STATE Create teacher network and student network;
		\FOR{\(t=1\) to \(100\)}
		\STATE Create current state information \(s_t\);
		\IF{\(\textit{val\_acc} > A_{\text{base}}\)}
		\STATE Clip ($\sigma_\theta$) by \text{clip} ($\sigma_\theta, 0.01, 0.1$);
		\ENDIF
		\STATE Sample action \(\alpha_t\) from \(\pi_\theta\) and update \(\tau_t\);
		\STATE Calculate reward \(R_t\) and clip reward by \(\text{clip}\);
		\STATE Update replay buffer \(\boldsymbol{B} \leftarrow \boldsymbol{B} \cup s_t\);
		\STATE Update \(\pi_\theta\) through \textbf{function} policy\_update;
		\STATE Update \(\mathbf{W}_S\) by Adam optimizer.
		\ENDFOR
	\end{algorithmic}
	\vspace{-0.2em}
\end{algorithm}

\begin{table}[!ht]
	\centering
	\caption{Comparison of model performance indicators}
	\label{tab:model_comparison}
	\renewcommand{\arraystretch}{0.7}
	\resizebox{\linewidth}{!}{
		\begin{tabular}{ccccc}
			\toprule
			\makecell{Model} &
			\makecell{Total\\Parameters} &
			\makecell{Accuracy} &
			\makecell{Inference\\Latency (ms)} &
			\makecell{RAM\\Usage (MB)} \\
			\midrule
			ResNet & 737,028 & 91.85\% & \textbf{2.74} & 3304.9 \\
			LMSC & 1,337,096 & 82.87\% & 2.78 & 3323.8 \\
			ShuffleNet-v2 & 935,384 & 94.73\% & 2.83 & 3179.8 \\
			Lite-HRNet-KD & \textbf{156,064} & \textbf{98.38}\% & \textbf{2.74} & \textbf{2444.2} \\
			\bottomrule
		\end{tabular}
	}
	\vspace{-1em}
\end{table}

\section{Conclusion}

In this paper, we proposed a dynamic KD-enabled wireless RFF-LLM framework via the PPO that adequately transfers the feature extraction capability of the RFF-LLM to the lightweight model Lite-HRNet. In a few iterations of the PPO, the temperature converged to a stable optimal range, which overcomes the local optimal dilemma in static distillation. Meanwhile, the experimental results on the self-built real-world dataset demonstrated that the dynamic-distilled Lite-HRNet achieved better performance in different indicators than the baseline methods.

\ifCLASSOPTIONcaptionsoff
  \newpage
\fi



\bibliographystyle{IEEEtran}
\bibliography{IEEEabrv,sigproc} 
\end{document}